\documentclass[10pt,journal,compsoc]{IEEEtran}

\IEEEoverridecommandlockouts
\usepackage{cite}
\usepackage[utf8]{inputenc}
\usepackage[super]{nth}
\usepackage{graphicx}

\usepackage[linesnumbered,ruled]{algorithm2e}

\usepackage{textcomp}
\usepackage{amssymb}
\usepackage{multirow}
\usepackage{verbatim, color}
\usepackage{subfigure}
\usepackage{amsmath}
\usepackage{listings}
\PassOptionsToPackage{table}{xcolor}
\usepackage{tabularx}
\usepackage{booktabs}
\usepackage{colortbl}
\usepackage{fancybox}
\usepackage{hyperref}

\makeatletter

\long\def\diffcolor#1#2\@nil{color\string#1diff}

\def\verbatim@processline{%
	\nointerlineskip\noindent\rlap{%
		\colorbox{\expandafter\diffcolor\next..\@nil}{%
			\the\verbatim@line}}\par}
\definecolor{color diff}{rgb}{1,1,1}
\definecolor{color-diff}{rgb}{1,.5,.5}
\definecolor{color+diff}{rgb}{.5,1,.5}
\newcommand{\hypobox}[1]{\begin{center}%	
		\noindent\thicklines\setlength{\fboxsep}{7pt}%	
		\cornersize{0}\Ovalbox
{\begin{minipage}{3.2in}%	
					\vspace{-0.01cm}
				\textit{#1}
					\vspace{-0.01cm}
\end{minipage}}
 \end{center}}

\def\BibTeX{{\rm B\kern-.05em{\sc i\kern-.025em b}\kern-.08em
		T\kern-.1667em\lower.7ex\hbox{E}\kern-.125emX}}
	
\definecolor{dkgreen}{rgb}{0,0.6,0}
\definecolor{gray}{rgb}{0.5,0.5,0.5}
\definecolor{mauve}{rgb}{0.58,0,0.82}

\newcommand{\tool}{A3}

\begin{document}
%used for listing DO NOT REMOVE
\lstset{frame=tb,
	language=Java,
	aboveskip=3mm,
	belowskip=3mm,
	showstringspaces=false,
	columns=flexible,
	basicstyle={\small\ttfamily},
	numbers=none,
	numberstyle=\tiny\color{gray},
	keywordstyle=\color{blue},
	commentstyle=\color{dkgreen},
	stringstyle=\color{mauve},
	breaklines=true,
	breakatwhitespace=true,
	tabsize=3,
	escapechar=\%
}
%end used for listing DO NOT REMOVE
	
		\title{~\tool: Assisting Android API Migrations Using Code Examples}

\author{Maxime~Lamothe, ~\IEEEmembership{Member,~IEEE,}
	Weiyi~Shang,~\IEEEmembership{Member,~IEEE,}
	Tse-Hsun~(Peter)~Chen
	
	\IEEEcompsocitemizethanks{\IEEEcompsocthanksitem Department of Computer Science and Software Engineering, Concordia University, Montreal,
		Canada.\protect\\
		E-mail: (max\_lam, shang, peterc)@encs.concordia.ca}

}

\markboth{Journal of \LaTeX\ Class Files,~Vol.~14, No.~8, August~2018}%
{Shell \MakeLowercase{\textit{et al.}}: Bare Demo of IEEEtran.cls for Computer Society Journals}

\IEEEtitleabstractindextext{%
	\begin{abstract}
		%Under Review
		The fast-paced evolution of Android APIs has posed a challenging task for Android app developers. To leverage Androids frequently released APIs, developers must often spend considerable effort on API migrations. Prior research and Android official documentation typically provide enough information to guide developers in identifying the API calls that must be migrated and the corresponding API calls in an updated version of Android (\emph{what} to migrate). However, API migration remains a challenging task since developers lack the knowledge of~\emph{how} to migrate the API calls. There exist code examples, such as Google Samples, that illustrate the usage of APIs. We posit that by analyzing the changes of API usage in code examples, we can learn API migration patterns to assist developers with API Migrations. 

In this paper, we propose an approach that learns API migration patterns from code examples, applies these patterns to the source code of Android apps for API migration, and presents the results to users as potential migration solutions. To evaluate our approach, we migrate API calls in open source Android apps by learning API migration patterns from code examples. We find that our approach can successfully learn API migration patterns and provide API migration assistance in 71 out of 80 cases. Our approach can either migrate API calls with little to no extra modifications needed or provide guidance to assist with the migrations. Through a user study, we find that adopting our approach can reduce the time spent on migrating APIs, on average, by 29\%. Moreover, our interviews with app developers highlight the benefits of our approach when seeking API migrations. Our approach demonstrates the value of leveraging the knowledge contained in software repositories to facilitate API migrations.

	\end{abstract}
	
	}

\maketitle

\IEEEdisplaynontitleabstractindextext

\IEEEpeerreviewmaketitle

\IEEEraisesectionheading{\section{Introduction}
\label{sec:intro}}
Software maintenance is one of the most expensive activities in the software development process~\cite{Glass:2001:FFF:624643.626281}. To reduce the maintenance cost, developers often rely on reusing available software. The reused software helps abstract the underlying implementation details and can be integrated into innumerable software projects. In particular, calling application programming interfaces (APIs) is a common software reuse technique used by developers~\cite{venkatesh_wang_zhang_zou_hassan_2016, sawant_bacchelli_2016}. These APIs provide well-defined programming interfaces that allow their users to obtain desired functionality without forfeiting development time.

However, in today's fast-paced development, APIs are evolving frequently. The Android API is one such example of a rapidly-evolving and widely used API~\cite{Bavota2015, mcdonnell_ray_kim_2013}. 
Prior studies~\cite{mcdonnell_ray_kim_2013, lamothe_shang_2018} found that Android is evolving at an average rate of 115 API updates per month. Such evolution may entail arbitrary release schedules and API deprecation durations and may involve removing functionality without prior warning~\cite{Sawant2017a}. Therefore, users must regularly study the changes to existing APIs and decide whether they need to migrate their code to adopt the changes. 
In fact, a prior study shows that developers are slower at migrating API calls than the API evolution speed itself~\cite{mcdonnell_ray_kim_2013}. As a consequence, there is fragmentation in the user base and slow adopters who miss out on new features and fixes~\cite{mcdonnell_ray_kim_2013}.

Existing migration recommendation techniques~\cite{wu_gueheneuc_antoniol_kim_2010, dagenais_robillard_2009, henkel_diwan_2005, nguyen_hilton_codoban_nguyen_mast_rademacher_nguyen_dig_2016} typically focus on identifying \emph{what} is the replacement of a deprecated API (e.g., one should now be using methodB instead of methodA), instead of \emph{how} to migrate the API calls for the replacement (e.g., how to change the existing code to call methodB). However, a recent experience report shows that all too often, Android API official documentation clearly states \emph{what} to replace for a deprecated API, while actually performing API migrations is still challenging and error prone~\cite{lamothe_shang_2018}.

There exist many publicly-available code examples online illustrating API usages. As an example, Google provides a set of sample Android projects on the Google Samples repository~\cite{Google_samples_cite}. Developers often study these sample projects and other code examples (e.g., code from open source Android apps) to help them with API migration~\cite{venkatesh_wang_zhang_zou_hassan_2016, treude_robillard_2016, Wang2014, Linares-Vasquez2014}. Nevertheless, studying the code examples to know the changes needed for API migrations is a manual and time-consuming process. Furthermore, identifying \textit{where} and \textit{how} to apply migration changes puts an extra burden on developers during software maintenance.

\begin{figure*}[tbh]
	\centering
	{\small
\begin{verbatim}
  view.setFocusableInTouchMode(true);
- view.setBackgroundColor(mFrag.getresources().getColor(R.color.default_background));
+ view.setBackgroundColor(mFrag.getresources().getColor(R.color.default_background, null));
  view.setTextColor(Color.WHITE);
\end{verbatim}%
	}
	\caption{An API migration example of the \textit{getColor} API, in the \textit{GridItemPresenter} class of \textit{androidtv-Leanback} project in commit~\textit{6a96ad5}.}
	\label{fig:example_migrationt}
\end{figure*}

In this paper, we propose an approach, named~\tool, that mines and leverages source code examples to assist developers with API migration. We focus on Android API migrations, due to Android's wide adoption and fast evolution~\cite{mcdonnell_ray_kim_2013}. Our approach automatically learns the API migration patterns from code examples taken from available code repositories, thereby providing varied example patterns. Afterwards, our approach matches the learned API migration patterns to the source code of the Android apps to identify API migration candidates. If migration candidates are identified, we apply the learned migration patterns to the source code of Android apps, and provide the resulting migration to developers as a potential migration solution.  

To evaluate our approach, first~\tool~learns Android API migration patterns from three sources of code examples: 1) official Android code examples provided by Google Samples~\cite{Google_samples_cite}, 2) migration patterns that are learned from the development history of open source Android projects i.e., FDroid~\cite{f-droid} and 3) API migration examples that are manually produced by users. Our approach then applies Android API migrations to open source Android apps from FDroid and we leverage their test suites and manually run the apps to validate the correctness of the migration. Furthermore, we compared our approach to LASE~\cite{Kim2013}, a tool meant to apply code edits learned from examples. Moreover, we carry out a user study to determine the actual and perceived usefulness of our approach. In particular, we answer three research questions.

\begin{description}

\item[\textbf{RQ1}] \textit{Can we identify API migration patterns from public code examples?}

Our approach can automatically identify 80 migration patterns with 96.7\% precision in Android APIs used in public code examples, and obtains a recall of 97\% using our seeded repository.

\item[\textbf{RQ2}] \textit{To what extent can our approach provide assistance when migrating APIs?}

Based on 80 migrations candidates in 32 open source apps, our approach can generate 14 faultless migrations, 21 migrations with minor code changes, and 36 migrations with useful guidance to developers. Furthermore, interviews with four developers highlight a positive developer response to our migration examples.

\item[\textbf{RQ3}] \textit{How much time can our approach save when migrating APIs?} 

Through a user study with 15 participants and 6 API migration examples, we show that our approach provides, on average, a 29\% migration time improvement and is seen as useful by developers.

\end{description}

Previous research has proposed approaches such as Sydit~\cite{Meng_2011} and Lase~\cite{Kim2013} to help developers with API migration; however, these approaches must be manually pointed towards pre-migrated examples without the ability to automatically retrieve or identify them~\cite{Robillard_book_2014}. Furthermore, the effectiveness of code examples on migration is affected by the context of the examples, whereby examples with closer contexts will waste less developer time when testing extraneous cases~\cite{Robillard_book_2014}. Therefore, by considering multiple examples from different contexts our approach generates well-fitted migration solutions.

Our approach can be adopted by Android app developers to reduce their API migration efforts to cope with the fast evolution of Android APIs. Our approach also exposes the value of learning the knowledge that resides in rich code examples to assist in the various tasks of API related software maintenance.

\textbf{Paper Organization.} Section~\ref{sec:background} provides a real-life example of an API migration to motivate this study. Section \ref{sec:setup} describes our automated approach,~\tool, that assists in Android API migration. Section~\ref{sec:experiment} presents the design of the experiments used to evaluate our approach and Section~\ref{sec:experiment_results} presents the results of our experiments. Section \ref{sec:related} provides a short survey of related work. Section~\ref{sec:threats} describes threats to the validity of this study. Finally, Section~\ref{sec:conclusion} concludes the paper.

\section{A Motivating Example}
\label{sec:background}

In this section, we present an example, which motivates our approach based on learning migration patterns from code examples to assist in API migration.

In Android API version 23, the \textit{Resources.getColor} API (as shown in Listing~\ref{listing:getColor_before_migration}) was deprecated and replaced (as shown in Listing~\ref{listing:getColor_after_migration}).  In fact, the deprecation and replacement (\emph{what} to migrate) are clearly shown in the official Android documentation~\cite{android_developers_get_color}. However, even with the help from the documentation, the addition of a new parameter provides new challenges to the Android app developers. In particular, developers may not have the knowledge necessary to retrieve the \emph{Theme} information nor to initialize a new object for the \emph{Theme} to make the new API call. Moreover, since the old API call does not require any \emph{Theme}, even if developers can provide a \emph{Theme}, there is no information on how to preserve backward compatibility.

On the other hand, there exists open source example projects on Google Samples~\cite{Google_samples_cite}, i.e., \textit{androidtv-Leanback}, a project on Github which presents several uses of the \textit{Resources.getColor} method. 
With the introduction of a new Android API version, these code examples are also updated. By looking at an example, we find that it clearly demonstrates how to call the \textit{Resources.getColor} API (see Figure~\ref{fig:example_migrationt}). From the changes to the code example, we can see that to maintain backward compatibility, developers can simply pass a \emph{null} object to the API. Without such an example, figuring out that a \textit{null} value preserves backward compatibility would require trial and error from developers. By learning such a migration pattern in the code examples, the effort of the challenging task of \emph{how} to migrate an API call can be reduced for developers.

\begin{lstlisting}[caption={Resources.getColor API before migration},label={listing:getColor_before_migration}]
public class Resources extends Object {
    public int getColor(int id) 
}
\end{lstlisting}

\begin{lstlisting}[caption={Resources.getColor API after migration},label={listing:getColor_after_migration}]
public class Resources extends Object {
    public int getColor(int id, %\textcolor{red}{Resources.Theme theme}) 
}
\end{lstlisting}

However, finding these migrations on Google Samples is a laborious endeavor. First of all, the code examples do not index their usage of APIs. Developers may need to search for the API usage of interest from a large amount of source code. Second, even with the API usage in the code examples, developers need to go through the code history to understand the API migration pattern, i.e., \textit{how} to apply these changes, which can be much more complex than the aforementioned example. Finally, even with the migration pattern, developers need to learn how to apply the migration on their own source code, which can be a challenging task~\cite{lamothe_shang_2018}.

Taking the aforementioned API \textit{Resources.getColor} as an example, we can detect a total of 1,626 places where the \textit{Resources.getColor} Android API is called in its deprecated form on a sample of 1,860 open source Android apps from FDroid~\cite{f-droid}. Migrating \textit{Resources.getColor} to Android API version 23 or later for all of those apps requires a significant amount of effort. Automating the above-described migration approach and providing the information of the learned pattern to developers can significantly reduce the required migration effort. 

In the next section, we present our approach, named~\tool, which provides assistance in Android API migration. 

\section{Approach}
\label{sec:setup}

\vspace{-0.2cm}	
\begin{figure*}[t]
	\vspace{-0.2cm}

	\subfigure[Step 1: Learning API migration patterns from code examples]{
		\includegraphics[width=\textwidth]{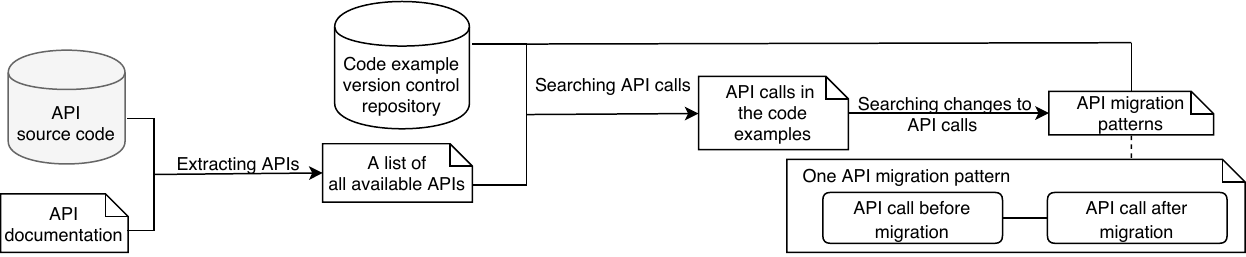}
	}
	\vspace{-0.2cm}	
	\subfigure[Step 2: Applying learned API migration patterns to API calls in the source code]{
		\includegraphics[width=\textwidth]{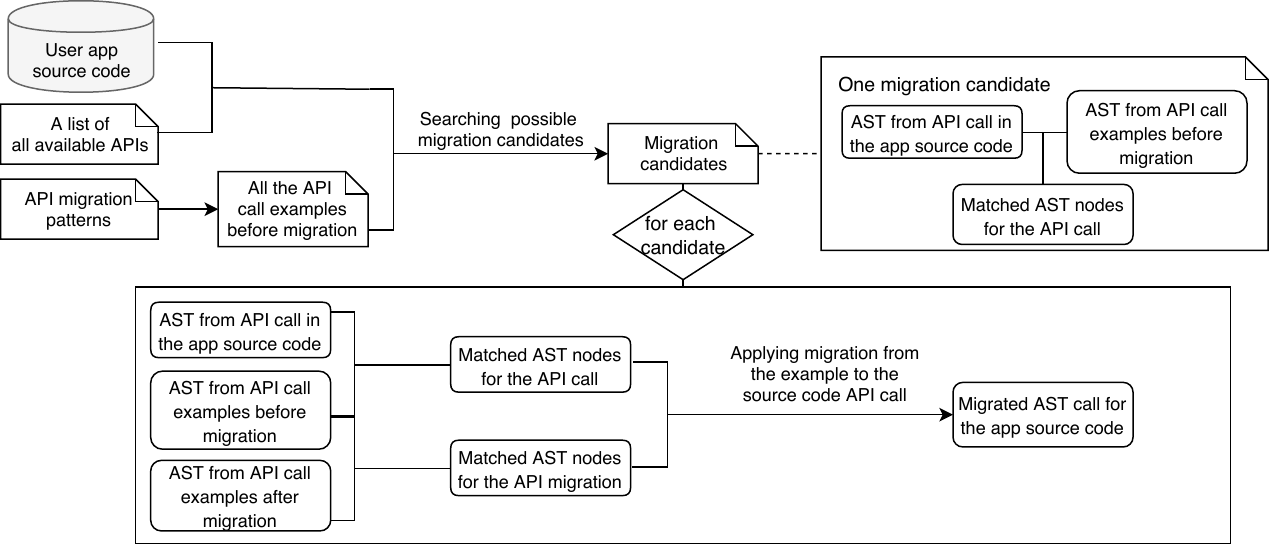}
	}

		\caption{An overview of our approach, named~\tool, that assists in API migration}
		
	\label{fig:methodology_migration}

\end{figure*}

In this section, we present an approach, \tool, that assists in API migration by learning how to migrate API calls from code examples. Our approach consists of two steps: 1) identifying API migration patterns from code examples and 2) applying the migration pattern that is learned from the example to the source code of the Android apps. An overview of our approach is shown in Figure~\ref{fig:methodology_migration} and an implementation of our approach is publicly available online\footnote{The repository which contains the tool and data presented in this paper can be found at: \url{https://github.com/LamotheMax/A3}.}.

\subsection{Learning API migration patterns from code examples }
\label{sec:step1}
In the first step of our approach, we mine readily available code examples to learn API migration patterns. Such code examples can be found through online code repositories such as GitHub and FDroid~\cite{f-droid}. 
Our approach also supports self-made examples, which allows users to produce their own migrated stubs, or use their own projects as data to feed forward into other projects.

\subsubsection{Extracting APIs}

The lack of uniformity between API documentation and source code presents a challenge to API migration approaches. There is currently no easy way to obtain a verified list of all migrated API methods for all versions of the Android API. Furthermore, the sheer size of the Android API presents a challenge to API migration. Manually searching source code for all deprecated, removed, and modified APIs is an arduous task.

Hence, we automatically extract all the Android APIs for every Android version that is available from the official Android API documentation. However, the Android documentation may have discrepancies to the APIs in their source code\footnote{We reported an issue in the Google issue tracking system for two discrepancies that we found between the Android API documentation and the source code repository.}. For every version of Android, we obtain the Android source code and parse the source code using Eclipse JDT AST parser~\cite{JDTAST}. We identify a list of all public methods as available APIs. 

\subsubsection{Searching API calls from code examples}

\begin{algorithm}	
	\SetKwInOut{Input}{Input}
	\SetKwInOut{Output}{Output}	
	\Input{Code example repository $exRepo$ and    
		available API list $apiList$}
	\Output{Migration example}
	$/*$ $First$ $do$ $basic$ $lexical$ $matching$ $*/$
	
	\ForEach{$api$ in $apiList$}{
		$apiCallData$ $\gets$ $api$.get($callData$)
		
		$apiParams$ $\gets$ $apiCallData$.getParamCount()
		
		$apiName$ $\gets$ $apiCallData$.getName()
		
		$potentialMigrations$ $\gets$ $null$
		
		\ForEach{$commit$ in $exRepo$}{
			$potentialMigrations$ $\gets$ $commit$.getAPIcalls($apiName$, $apiParams$)
		}
		
		$/*$ $Selective$ $check$ $for$ $all$ $potential$ $migrations$ $*/$
		
		\ForEach{$call$ in $potentialMigrations$}{
			$callAST$ $\gets$ $call$.buildAST()
			
			\uIf{$callAST$.matches($apiCallData$)}{saveExample($call$)}
		}	
	}
	\caption{Our algorithm for searching API migration examples.}
	\label{alg:searchMigration}
\end{algorithm}

In this step, we identify the API calls that are available in the code examples. A pseudocode algorithm of this step is presented in Algorithm~\ref{alg:searchMigration}. One may design an approach that builds AST for all available code examples as well as the targeted Android app source code. Afterwards, one may use the ASTs to match the API calls in the source code to the code examples. However, such an approach would be time consuming due to the fact that 1) building ASTs is a time-consuming endeavor, 2) the complexity of AST matching is non-negligible and 3) the number of Android APIs that are available is large. 

In order to reduce the time needed to identify API calls, and thereby reduce the challenges faced by developers who seek to migrate APIs, we first use basic lexical matching to limit the scope of needed AST building and matching. In particular, we first search all the files in the code example for strings that match API names. This technique allows us to quickly get a preliminary list of potential API calls. These files are then selected for further processing as potential matches. Although the basic lexical matching can lead to false positives, the goal is merely scoping down the search space and the following AST matching can remove these false positives. For example, in Figure~\ref{fig:example_migrationt} we specifically search for the keyword~\emph{getColor}, although it is possible for this basic search to falsely identify a similar post-migration API as a migration target, once the refined check with AST matching is complete i.e. these \emph{getColor} false positives would be filtered out if they did not have exactly one integer parameter, the right return type, and all other discernible AST properties for~\emph{getColor}.

After the basic lexical matching, we apply AST matching on the files with potential matches. This allows us to scope down the previously identified results to obtain high precision findings without sacrificing too much performance. We leverage JDT~\cite{JDTAST} to parse the source code in the code examples to generate their corresponding ASTs. For each method invocation in the AST, we compare with the APIs that are potentially matched from the lexical search. If the AST can be fully built, we aim to obtain the perfect matches between method invocation and API declaration. In particular, a perfect match requires matching the method invocation name to an API declaration as well as having perfectly matched types for all parameters. In some cases, the code examples may not be fully complete (e.g., missing some external source code files or dependencies), leading to partially built ASTs. With the partially built file level ASTs, we consider a match if there exist the correct import statement of the API, the correct method name, and the correct number of parameters. All method instances are saved, along with their invocation string. This data can quickly be reviewed by a developer to determine if a false positive was detected, making our approach transparent and understandable. Developers interviewed in RQ2 presented in Section~\ref{sec:experiment_results}, were presented and reviewed the migration instances that were related to their familiar app and confirmed our results.

\subsubsection{Learning API migration patterns}

In this step, we search API calls for every version of the code examples. For instance, if code examples such as Google Samples~\cite{Google_samples_cite}, or FDroid projects~\cite{f-droid} are hosted in version control repositories, we detect API calls in every commit of the repository. 

Searching every commit for a multitude of projects requires surmounting the challenge imposed by the large scale of available data from API user projects. We surmount the challenge imposed by the large amount of data available by leveraging the basic \emph{diff} in the version control system (like Git) to determine in which commits a specific string is modified to collect commits that contain changes to API calls.

Commit level migrations allow us to reduce the amount of code modifications that could obfuscate an API migration. If we reduce the search scope to a granularity coarser than commit level we might miss certain migrations.
Similar to the lexical search from the last step, the use of a diff tool is merely scoping down our search space. For the commits that potentially impact API calls, we parse the AST of the changed files in the commit. We compare the ASTs generated from the source code before and after the commit. If the API invocations in the AST are modified in the commit, we consider the commit as a potential API migration pattern. Hence, each API migration pattern consists of an AST built from the example before the migration and an AST built after the migration.  

For the code examples that are hosted as text files outside of repositories, we apply a text diff on each two consecutive versions of the example instead of using commits to scope down the search space. The secondary AST matching step on these text files is identical to that from the version control repositories.

\subsection{Applying learned API migration patterns to API calls in the source code}
\label{sec:step2_approach}
In our second step, we collect all the API migration patterns (i.e., ASTs built from the example before and after the migration) from the first step and try to apply these patterns to the API calls in the Android app source code. 

\subsubsection{Searching possible migration candidates}

Similar to our first step, to resolve the challenges imposed by the scale of the problem at hand, we reduce the search space, our approach first uses API names to lexically search for API calls with available migration patterns in the source code of the targeted Android app. 

Since migrations can be dependent on the context of surrounding code, we cannot assume that one migration example will suit all possible use cases. For example, API such as \emph{Resouces.getColor(int)} can be present in a variety of use cases. To match valid migration patterns, we must not only match API calls, but also determine if an example matches a user's usage of the API. We leverage data-flow graphs to match the API calls in the migration patterns and the targeted Android app sources. We construct a data-flow graph from the API call example ``before'' the migration in the migration patterns. We also construct data-flow graphs in the API calls in the Android app source code. Only if the data-flow graph from the example is a subgraph of a potential API call in the Android app source code do we then consider this a migration candidate. This allows us to assume that the example API being used as a migration pattern has a similar use case to the API call in the targeted Android app.

\subsubsection{Applying the migration from the example to API calls in the source code}

\begin{algorithm}

	\SetKwInOut{Input}{Input}
	\SetKwInOut{Output}{Output}
	
	\Input{Migration mapping $mappedDFG$ and    
		client data-flow graphs $clientDFG$}
	\Output{Migrated data-flow graph}
	
	$/*$ $Traverse$ $all$ $data-flow$ $graphs$ $*/$
	
	\ForEach{DFG in $clientDFG$}{
		$DFGMap$ $\gets$ $mappedDFG$.get($DFG$)
		
		$changedAPIs$ $\gets$ $DFGMap$.getChangedAPI()
		
		$/*$ $Migrate$ $all$ $migrateable$ $APIs$ $*/$	
		
		\ForEach{changedAPI in changedAPIs}{
			changedNodes $\gets$ $DFGMap$.getDataLinks($changedAPI$)
			
			missingNodes $\gets$ $DFGMap$.getNodesToAdd($changedAPI$)

			$/*$ $Adjust$ $the$ $data-flow$ $graph$ $*/$
			
			$DFG$.addNewNodes(missingNodes)
			
			$DFG$.migrateDFG(migrateableNodes)}
	}
	
	\caption{Our algorithm for migrating code.}
		\label{alg:migrationCode}
\end{algorithm}

Existing API migrations can contain implementation details that cause incompatibilities with other projects. Therefore, our approach must mitigate the challenges imposed by implementation details and allow developer interaction. We employ data-flow graphs, rely on the large number of migration examples provided by open-source projects to obtain our migration examples, and let the developer have final say at every step of the approach.

If any migration candidate is found, we then attempt the migration. We first compute the migration mappings by comparing the examples before and after API migration. This mapping contains any changes that must be made to existing code statements, obtained by comparing the names and types of each code statement in the data-flow graph. The migration mapping also contains any new code statements that are present in the ``after'' migration example but were missing in the ``before'' migration example.

In order to obtain an accurate migration mapping between the ``before'' and ``after'' examples (i.e., changes that were made to migrate the API), we need to eliminate the changes on AST that are not related to the API migration. We achieve this by relying on the data-flow graphs that are built from the examples. We first remove all the nodes in the data-flow graph where all the associated nodes are perfectly matched between the ``before'' and ``after'' examples. Since they are perfect copies of one another, those nodes cannot contribute to a migration.  We keep the nodes in the data-flow graphs that remain unmatched and are associated with the node that is of interest to the API call. Finally, we compare the nodes that are kept to find the matched data-flow graph for the API call in the Android app after migration.

Once we obtain the most likely migrated data-flow graph in the ``after'' API migration example, we produce a backward slice of the data-flow graph starting from the API call. In other words, we only look at nodes that give data to the API call. Based on our sliced data-flow graph, we then map each node in the ``before'' example to the most closely matched node in the ``after'' example. Any unmatched data linked nodes are considered to be new nodes and are saved to be added during migration.

Finally, we use the migration mapping to transform the project source code into the ``after'' API migration example. Pseudocode of our migration algorithm can be found in Algorithm~\ref{alg:migrationCode}. The transformation also looks for any object types that are matches between the ``before'' example and the project source code to infer the names used in the Android app source code, to prevents the introduction of new variable names. Our approach can produce migrations that are interprocedural and intraprocedural. However, we limit the migration scope of our approach to single files since prior work has found that field and method changes account for more than 80\% of migration cases~\cite{Cossette_bradley_walker_robert_2012}. Adding migration across multiple files would increase the chance of making mistakes, both when mining and applying migration patterns, and only account for a minority of migration cases.

\begin{figure*}[tbh]
	\centering
	\includegraphics[width=\textwidth]{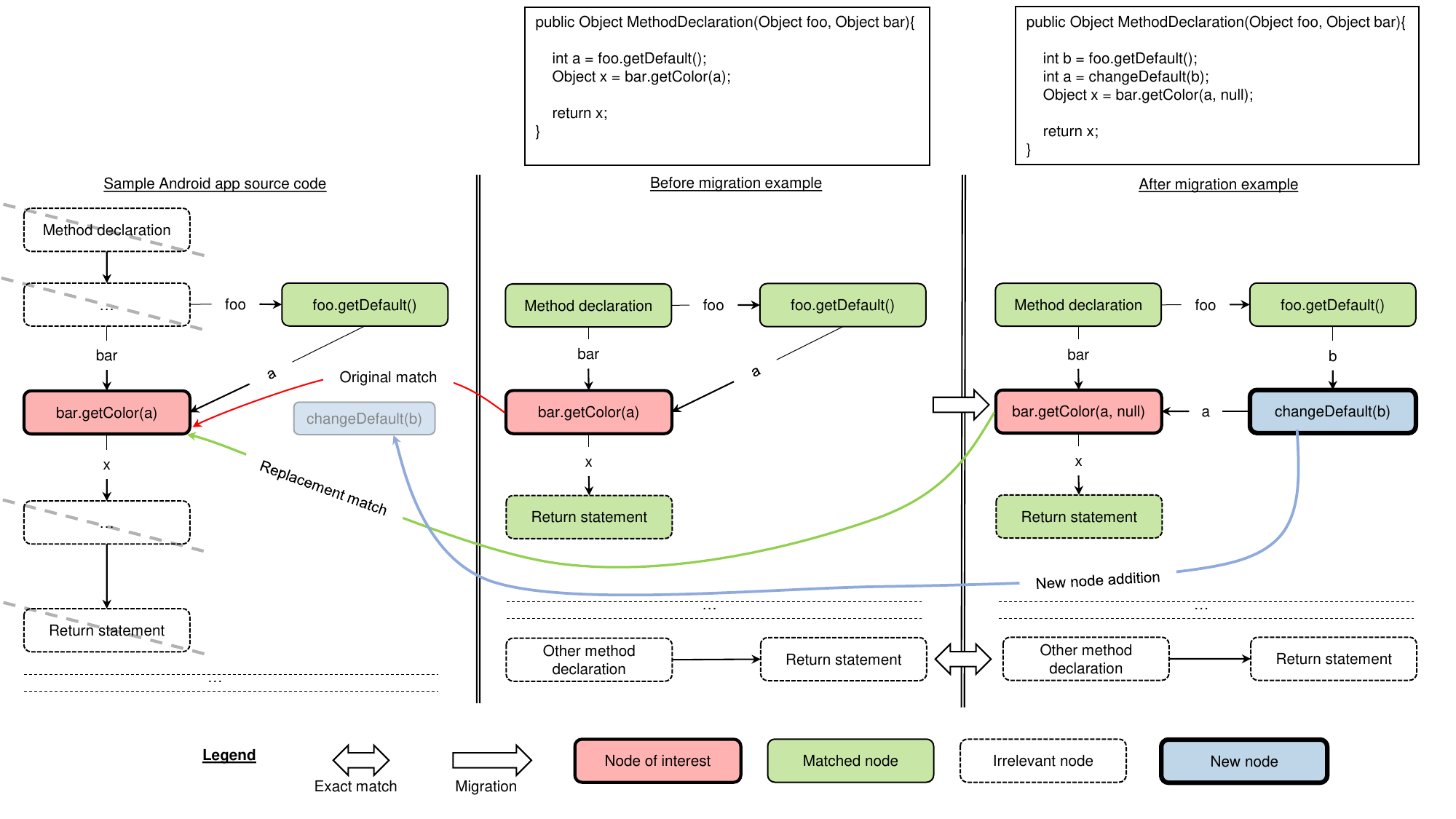}
	\caption{Example of applying a migration from an example to Android app source code}
	\label{fig:migration_example}
\end{figure*}

To better explain our approach, Figure~\ref{fig:migration_example} illustrates an example of applying a migration from an example to an API call. The \emph{before} and \emph{after} migration examples illustrate potential example code obtained from sample projects. We can see that methods which are exact matches are seen as irrelevant since they add no information to the migration. Similarly, we can see that nodes after the migration node of interest, in this case a return statement, are also seen as irrelevant nodes. This is because it does not matter what the user does with their code after the API call (i.e., not related to the usage of the API). Other nodes are then matched between \emph{before} and \emph{after} examples to determine which node is most likely the migration. Any nodes that do not obtain a match are considered new nodes required for migration and will be added in user projects, as denoted by the color blue in our example. Once a match is produced between migration examples, the migration mapping can be applied to the user example, as presented by the arrows in Figure~\ref{fig:migration_example} and as explained by the pseudocode in Algorithm~\ref{alg:migrationCode}.

\section{Evaluation design}
\label{sec:experiment}
To determine the extent to which our approach,~\tool, can be used to assist with API migration, we conducted a series of experiments. We first present the research questions which we use to evaluate our approach. Then, we present our data acquisition approach for Android apps and code examples.  

\subsection{Research questions}

We aim to evaluate our approach by answering three research questions. We present our research questions and their motivations in the rest of this subsection.

\noindent \textbf{RQ1} \textit{Can we identify API migration patterns from public code examples?} 

Our approach uses code examples to automatically learn API migration patterns. Therefore, the availability of migration code examples is very important to our approach. In this RQ, we study how many API migration patterns we can find in publicly available code examples and calculate the \textit{precision} of our approach. We also engaged the help of 20 volunteer developers to produce a sample migration dataset in order to obtain a known number of migration examples and determine the \textit{recall} of our approach.

\noindent \textbf{RQ2}  \textit{To what extent can our approach provide assistance when migrating APIs?}

In this RQ, we use the publicly available and user produced examples to automatically learn and apply migration patterns. We evaluate and quantify the extent to which our approach,~\tool, can leverage these examples to assist developers in migrating API calls.

\noindent \textbf{RQ3}  \textit{RQ3: How much time can our approach save when migrating APIs?}

In this RQ, we enlisted 15 developers to participate in a user study to evaluate the time savings provided by~\tool. We timed the participants while they migrated an amalgamation of six API examples, for which they randomly received help from~\tool. The participants were also asked to rate the usefulness of our approach.

\subsection{Data acquisition}

In this subsection, we present our data acquisition approaches for the Android apps that we want to migrate and the sources of our code examples.

\subsubsection{Android apps for migration}

We selected our Android apps through the free and open source repository of Android apps: FDroid~\cite{f-droid}. We clone all the git repositories of FDroid apps that are hosted on GitHub and implemented in Java. In total, we obtained 1,860 apps. From these apps, we would like to perform API migration on the ones that are still actively under development, since they are more likely to benefit from migrating to the most recent Android APIs. Therefore, we only selected apps that had code committed in the six months prior to our study. In order to assist in later verifying the success of the migration, we selected only the apps that contained readily available tests. Furthermore, we only selected the apps that could be built with the official Android build system, Gradle, so that we could verify the functionality of apps after migration. We focused the evaluation of our approach for API migration on the 10 latest versions of the Android API since they account for 95.6\% of Android devices in the world~\cite{android_developers_dashboard}. 
Therefore, we selected the apps that target the 10 latest versions (API versions 19-28) of the Android API. This allowed us to collect a sample of 164 FDroid apps on which to test our migrations. All the other apps (1,696) were used to extract API migration examples.

\subsubsection{Sources of public code examples}

Our approach relies on code examples to learn API migration patterns. In the evaluation of our approach, we focus on two sources of public code examples: 1) official Android examples, i.e., Google Samples~\cite{Google_samples_cite} and 2) FDroid app development history. We also use Android API usage patterns extracted from our FDroid sample to construct the original examples given to our user study participants.

\noindent \textbf{Google Samples. }The Android development team provides code samples to assist app developers understand various use cases of the Android APIs. This code sample repository, aptly named Google Samples, contains 234 sample projects, 181 of which are classified as Java projects~\cite{Google_samples_cite}. We mined the Google Samples repository, hosted on GitHub, for Android API migration examples.

\noindent \textbf{FDroid app development history.} Due to the widely available open source projects, if one open source project migrates a deprecated API in their source code, other developers may leverage that migration as an example. With the publicly available development history of FDroid, we can leverage the API migrations that exists in the FDroid apps as code examples.

\subsubsection{Source of private code examples}
\label{private_examples}

Since it is impossible to determine how many migrations have been produced and can therefore be recovered from mining public repositories, we produced a sample migration dataset with the help of volunteer developers. This dataset was used in RQ1 to determine the recall of our approach when mining for migration examples.

We developed a dataset with the help of 20 participants (14 graduate students and six professional developers) to create migration examples for Android API methods that had known migrations. Based on our study of the FDroid repositories, we selected the 30 Android APIs which were most frequently called but did not have readily available public examples of migration in FDroid apps. We manually created stub classes based on multiple examples of existing code taken from FDroid apps to reduce the amount of code our volunteers had to read. We then presented these classes to our participants as files sampled from real apps that needed migration. The participants were each given five files, selected at random from our pool of 30 Android APIs that needed migration, in order to guarantee multiple samples of successful migrations for each API. Therefore, each API had at least three participants attempt the migration. The participants were told which API to migrate for each file, and were allowed to use the official documentation or any other source of information they deemed necessary to create suitable migrations. We manually verified the results of the participants and randomly selected working examples to seed a mine-able sample repository of 30 real migration code examples that were created by real developers and sampled from real projects~\footnote{The 30 code examples, and their migrations, are available as part of our replication package}.

\section{Evaluation results}
\label{sec:experiment_results}
In this section, we present the results of our evaluation by answering three research questions.

\noindent\textbf{RQ1: Can we identify API migration patterns from public code examples?}
\label{sec:rq1}

We focus on two sources of examples in this RQ: 1) official Android examples, i.e., Google Samples~\cite{Google_samples_cite} and 2) FDroid app development history. We measure the prevalence of API migration examples by determining the number of API migration examples that we can mine from our example sources. We only consider the APIs with migrations that have officially been documented by Android developers to validate our results.

In the 10 latest versions of Android (versions 19-28), we were able to manually identify 262 APIs which had documented migrations. Out of all the Android APIs with migrations, only 125 of those occurred in our sources of examples. 

\begin{table}[htbp]
	\caption{API migrations identified}
	\centering
	
	\begin{tabular}{lc}
		Data sources & \multicolumn{1}{r}{API identified} \\
		\hline
		FDroid API migrations & 79\\
		Google Samples API migrations & 10 \\
		\hline
		Total distinct API migrations found  & 80 \\
		Total API uses found in apps & 125 \\
		Total possible & 262 \\
	\end{tabular}%
	\label{tab:migrations_studied}%	
\end{table}%

By searching for API migrations (c.f. Step 1 in our approach in Section~\ref{sec:step1}) in the Google Samples~\cite{Google_samples_cite} and the development history of FDroid apps, our approach can automatically identify 10 and 79 API migration examples out of 125 API occurrences, respectively. Among those migration examples from Google Samples, only one API is not covered in the FDroid examples, giving us a total of 80 distinct API migrations found (see Table~\ref{tab:migrations_studied}). However, we note that the nine other examples from Google Samples that overlap with other sources are still useful due to the possible minor variations in ASTs that can occur when migrating. Some examples may provide a more suitable migration with less needed human effort from developers (c.f. RQ2 \& RQ3). We manually analyzed all examples mined and determined that our approach was able to find migration patterns with 100\%  precision in Google Samples (i.e., 10/10 examples correctly identified), and 96.3\% precision in FDroid apps (i.e., 79/82 examples correctly identified), giving a total precision of 96.7\%. Some examples are migration false positives due to user modifications that can be mistaken an migrations, we discuss examples that are not migrations in detail in RQ2.

To calculate the recall of our approach, we used the manually created sample of 30 Android migration examples that was created by volunteer developers (c.f. Section~\ref{private_examples}). Our approach was able to successfully extract 29 out of 30 migration examples from the seeded repository, giving us a recall of 97\%. The example which could not be extracted was due to line-breaks within the API call. It would be possible to fix this corner case either by preprocessing all commits to remove line-breaks, or modify our API search to allow for arbitrary line-breaks.

\begin{table}[hb!]
	\caption{Classified API migration patterns identified by our approach}
	\label{tab:classification}
	\begin{tabular}{c|l|r}
		\multicolumn{1}{l|}{Migration difficulty} & Type & \multicolumn{1}{l}{Frequency} \\ \hline
		\multirow{4}{*}{EASY} & Encapsulate & 6 \\
		& Move method & 4 \\
		& Remove parameter & 1 \\
		& Rename & 7 \\ \hline
		\multirow{2}{*}{MEDIUM} & Consolidate & 7 \\
		& Expose implementation & 10 \\ \hline
		\multirow{3}{*}{HARD} & Add contextual data & 23 \\
		& Change type & 15 \\
		& Replaced by external API & 7
	\end{tabular}
\end{table}

We manually classified the 80 distinct migration patterns identified by our approach according to pre-established API migration classifications~\cite{Cossette_bradley_walker_robert_2012}. Table~\ref{tab:classification} presents our classified patterns. As can be seen in Table~\ref*{tab:classification}, the majority (45/80) of the migrations we identified are considered hard to automate~\cite{Cossette_bradley_walker_robert_2012}. This implies that our approach can surmount hard technical challenges, as well as find easy-to-migrate examples. The easy difficulty migration implies that the migration patterns can normally be completed by most IDEs such as Eclipse~\cite{Cossette_bradley_walker_robert_2012}. For example, in API version 26, \textit{Notification.Builder.setPriority} was moved and renamed to \textit{NotificationChannel.setImportance} and \textit{CookieSyncManager.startSync} is no longer useful since it was encapsulated by the WebView class in API version 21 and is now automatically done by that class. Medium difficulty migrations (i.e., migrations previously considered partially automatable) were also found. We identified examples of API consolidation such as \textit{NetworkInfo.getTypeName} which was consolidated into the \textit{NetworkCapabilities} class in API version 28 and must now instantiate that class and call a different method \textit{hasTransport} new with constants. We also identified cases where implementation was exposed to give more control to the API users. For example the \textit{Notification.Builder.setDefaults} method started allowing users to enable vibration, enable lights and set sound separately in API version 26 rather than be done as part of one method, as it was done before API version 26. This means that our approach must be able to find code to instantiate all of these new functionalities. As previously mentioned, our approach also identifies hard to automate migration patterns. We identify patterns such as added contextual data in methods like \textit{Hmtl.toHtml} where new parameters were added in API version 24 to allow for more options and user control. In most of these cases a default value for new parameters is allowed, which our approach can mine from existing projects to allow users to migrate their project. Changed types such as faced when migrating to \textit{Message.getSenderPerson} from \textit{Message.getSender}, in API version 28, can be handled by our approach by following the control flow graph of calls and modifying the calls appropriately based on previous migrations. Finally, APIs replaced by external APIs such as the FloatMath APIs in API version 22 \textit{cos, sin, sqrt} can be changed to their new library format, and the import statement can be created. The user then only has to make sure that the new API library is included in their project.

\hypobox{Our approach can automatically identify 80 migration patterns with 96.7\% precision in Android APIs used in public code examples, and obtains a recall of 97\% using our seeded repository.}

\noindent\textbf{RQ2: To what extent can our approach provide assistance when migrating APIs?}
\label{sec:rq2}

In this RQ, we examine whether our approach can provide assistance to API migration in Android apps. In particular, we apply our approach to migrate 80 API calls in 32 FDroid apps (based on both the migration patterns learned from public code examples and user generated examples). In order to examine whether a migration is successful, we leverage the tests that are already available in the apps and collect the ones that can exercise the migrated API calls. In order to avoid tests that are already failing before the migration, we run all the tests before the migration of the apps and only keep apps with passing builds. The apps that we selected had an average of 10 tests, with a minimum of 6 tests and a maximum of 52. Afterwards, we try to build the migrated apps. For the apps that can be successfully built, we run the collected tests again to check whether the migration is completed successfully. Furthermore, we also manually install and run the migrated apps to ascertain that the migration does not cause the app to crash. Furthermore, we provide comparisons with prior approaches such as LASE.

Previous studies have shown that producing exact automated migrations is a difficult task~\cite{Cossette_bradley_walker_robert_2012}. We consider this fact and attempt to mitigate it through the use of our approach. However, in cases where it is possible to present an exact migration to a user, such a task should be attempted to save development time. Therefore, in the best case scenario we attempt to provide an exact and automatic migration to users. Table~\ref{tab:migration_results} summarizes the results of our migrations. Our approach can provide assistance in 71 out of 80 migrations through faultless migrations, migrations with minor modifications, and through the examples we suggest when we experience unmatched guidance or unsupported cases. The following paragraphs discuss each different type of result in detail.

\begin{table}[tbh]
	\centering
	\caption{Automated migration results based on migration patterns learned from three different sources}
	\begin{tabular}{l|rrr}
		& \multicolumn{1}{c}{FDroid} & \multicolumn{1}{c}{\begin{tabular}[c]{@{}c@{}}Google\\ Samples\end{tabular}} & \multicolumn{1}{c}{\begin{tabular}[c]{@{}c@{}}Volunteer\\ developer\\ sample\end{tabular}} \\ \hline

		Faultless migration & 5 & 0 & 9 \\
		Migrated with minor mod. & 5 & 4 & 12 \\
%		Avg. tokens refactored & 3.3   & 4     & 4.25 \\
		Unmatched guidance & 18 & 1 & 9 \\
		False positive & 0 & 0 & 3 \\	
		Ex. was not a migration& 6 & 0 & 0 \\
		Unsupported cases & 7 & 0 & 1 \\
		\hline
		\textit{Total migrations} & \textit{41} & \textit{5} & \textit{34} \\
		\textit{Distinct API} & \textit{21} & \textit{4} & \textit{15} \\
	\end{tabular}%
	\label{tab:migration_results}%
\end{table}%

\noindent \textbf{Faultless migration.}
If all the tests are passed, and the app runs, without any further code modification, we consider the migration as exact. We succeeded at giving users an exact migration in 14 cases. Nine of these migrations were learned from the manually produced code examples from the examples manually created by participants. However, we were able to use five examples from FDroid app development history to produce faultless migrations. Such a result tells us that given a rich example, it is possible to provide fully automated working migrations.

\noindent \textbf{Migration with minor modifications.}
In cases where an app, after the migration, does not build and run, or does not pass the tests, we manually checked the error message. In such cases, the migration pattern may be correct, yet our automatically generated migration may need minor modifications to build and run the app and pass the tests. For each case, we determined a way for the migration to succeed with minimal code modification. An example of such modification can be adjusting a variable name. We were able to provide 21 migrations with minimal modifications. We consider the number of tokens that must be changed for the migration to be successful. The number of tokens changed was determined by post-modifications performed by the first author, who manually went through failed automated migration cases, modified the app’s code to make the migration successful and measured the size of the modification. We use the absolute number of tokens changed rather than a percentage of tokens matched since the automatically matched tokens do not require any effort.
We consider any modifications to a code token as a token modification. 

We found that the modifications needed for an imperfect but successful migration requires modifying between one to seven tokens (3.65 tokens on average). We consider the amount of effort needed on such modifications rather small, especially since they are mostly simple renames and the addition or removal of keywords. The brunt of the work, namely finding a migration candidate, finding a migration pattern, and matching this pattern, is provided by our approach,~\tool. Therefore, the user can simply ``glue'' any unattached pieces of example code into their application.

We examined the different scenarios that require minor modification to qualitatively understand the effort needed for such modifications. 
In total, we identify four reasons for such modifications: \emph{variable renaming}, \emph{missing the keyword \textbf{this}}, \emph{wrongly erasing casting}, and \emph{removing API calls}. 

As a design choice, we opt to conservatively not remove any API calls for migration, since mistakenly removing API calls may cause large negative impact to developers. Instead, our approach tells the users that a change must be made to the API, and the API call must then be manually removed. Although this may require the manual modification of removing several tokens, the effort of the change is minimal. We experienced three \emph{removing API call} cases. For example the~\emph{View.setDrawingCacheQuality(int)} method was made obsolete in API 28 due to hardware-accelerated rendering~\cite{android_developers_index}. This means that old code referring to drawing cache quality can be simply deleted, as the OS now handles this through hardware.

\noindent \textbf{Unmatched guidance.} It is possible for our approach to fail to match an example to a known migration (see Section~\ref{sec:step2_approach}-1). These cases exist due to the nature of our example matching. Since we consider the data-flow surrounding a migration, our examples must contain a similar data-flow graph to the Android app considered for migration. We conservatively opt for such an approach to reduce the number of falsely generated migrations, since they may introduce more harm to developers than assistance. As a result, if the Android app instantiates their API call in a different way than our examples, our approach will not attempt to automate the migration. For example, if the user uses nested method calls inside an API call and we cannot reliably map the return types of the nested method calls, we will not consider the code to match. Examples that contain unmatched but similar API migrations will be presented to the user, and they can choose the correct migration and manually apply it afterwards. 

For the 28 migrations for which none of our examples contained a match, 18 of these are from FDroid development history. This implies that the FDroid API usage is more often tailored for the apps' needs and is not often coded in a general manner. Future research may investigate automatically generalizing the usage of API to address this issue.

\noindent \textbf{False positives.} We consider a migration to be a false positive when our approach presents an unnecessary migration. This would occur if a migration example were erroneously matched to a non-migrateable method invocation in Android app code. This only occurred three times in our tests and the occurrences were all from the manually produced examples. We believe this to be the case due to the simplicity of the manually produced examples. Since the examples are simplified and contain very little context code, the corresponding data-flow graphs are often simple. This allows the examples to be used in a wider range of situations than the mined code samples. However, this leads to false positives as a trade-off, especially if method names are commonly used and few parameters are used (e.g. \emph{setContent()}). These false positives can be caught at compile time since they would not compile, and can therefore be corrected or discarded by the developer without harm to the app.

\noindent \textbf{Example was not a migration.}
As previously mentioned, since we obtain our migration examples through automated tool assistance (c.f. RQ1), it is possible for our approach to present faulty migration examples. This normally occurs when a deprecated API call is modified, but not migrated to the updated version of the API. An example of such a modification would be to rename the variables in the API call. This change would be mistakenly identified by our approach as a modification to an API call. Therefore, if such an example were to be used to attempt a migration, we would present a migration with no effect, and therefore cause no harm to the Android app. We experienced six such instances in the FDroid examples. However, we did not encounter these in our Google Samples nor the manually produced examples.

\noindent \textbf{Unsupported.}
There exist a few unsupported corner cases that our approach would not support. For example, we cannot automatically produce migrations if the API call is spread across a try catch block, or within a loop declaration or conditional statement declaration. As with unmatched guidance cases, we provide the user with migration examples that may be relevant to their migration, instead of attempting to provide automated migration on their code, therefore no harm is done to the code. We experienced eight such cases.

\begin{table}[tbh]
	\centering
	\caption{Comparison with LASE~\cite{Kim2013}}
	\label{tab:LASE_comparison}
	\begin{tabular}{l|rr}
		& \multicolumn{1}{l}{\textbf{A3}} & \multicolumn{1}{l}{\textbf{LASE}} \\ \hline
		Total Migrations Possible & \multicolumn{2}{c}{\textbf{17}} \\ 
		\hline
		Migration map successfully created & 17 & 15 \\
		Migration point successfully found & 9 & 1 \\
		Migration faultlessly applied & 7 & 1
	\end{tabular}
\end{table}

\noindent \textbf{Comparison with other approaches.}
To the best of our knowledge, there currently exists no other approach that can automatically identify API migration examples and automatically suggest and apply them to app code. Tools such as SEMDIFF~\cite{dagenais_robillard_2009}, and AURA~\cite{wu_gueheneuc_antoniol_kim_2010} find migration locations, but do not provide example-based automatic migrations. We find 97\% of migration locations based on our examples, which is close to the upper bound of what these approaches can achieve. We selected LASE~\cite{Kim2013} as a more direct comparison for our approach since LASE resembles the automatic migration of our approach while needing to be provided compilable examples. Due to LASE's need for compilable examples, we had to use~\tool~to first obtain migration examples to feed into LASE. Furthermore, LASE does not automatically determine a match between examples and must be manually pointed to paired examples through manually edited configuration files, which we had to produce for each example set. Other similar approaches are discussed further in Section~\ref{sec:related}. Due to the high degree of manual effort when manually producing configuration files for LASE, we chose a subset of 10 of the 32 apps used for RQ2 to compare with LASE. The apps were selected at random, and contained a total of 17 migration possibilities comprised of 13 distinct APIs. The results can be seen in Table~\ref{tab:LASE_comparison}.

LASE can build a refactoring map for 15 (88\%) migrations compared to 17/17 for~\tool. However, only 1/17 migrations were automatically mapped to migration points in the apps and subsequently applied to the apps. LASE appears to be highly dependent on the quality of the example and their similarity to the app code when compared to~\tool. We also find that LASE highly depends on complete matches in method AST. For example, if app code is inside an \emph{if} statement or \emph{try} block when the example code is not, if app code is instantiated differently than example code, or part of a method argument chain is in a different order to the example code, the AST sub-trees of the code will appear to be different. The differences in the sub-trees may lead to mis-detections between the primary example and the app code by LASE. We believe that since LASE was not designed for our use cases, it is intuitive that LASE is not optimal for our API migration task. 
On the other hand, ~\tool~is designed to mine examples from the online code example, and hence is less sensitive to factors such as as complete AST-alignment and therefore can more often provide migration links when examples are loosely aligned.

We conducted four semi-structured interviews concerning three open source apps from FDroid (\textit{Antennapod, K9Mail, and Wikipedia}). We contacted app developers that were contributors to the apps. Four developers offered to be interviewed for our study. Two developers for the Wikipedia app, one for Antennapod, and one for K9Mail. All of the developers had at least two years of experience in programming and Android app development. We presented each developer with migration candidates from the app they were familiar with. We also allowed them to search the web, and ask any clarifying questions if they had any. We then provided one example of a migration of the same API call done in another app, mined by our approach. 	
We asked four developers the following questions: 

\begin{enumerate}
	\item ``Do you think our approach correctly identified the migration candidate?''
	\item ``How confident would you be when migrating this API?''
	\item ``How do you feel about the examples provided by our approach to help you migrate?'' 
	\item ``Do you think there are limitations or potential improvements to our approach?''
\end{enumerate}

In all cases, the developers said that we had indeed found useful migration candidates. The migrations were judged to be easy, but time consuming. In all cases the developers said that our examples were \textit{``extremely useful''} or \textit{``very very useful''}, and that they would reduce the time to complete a migration from \textit{``a few minutes, down to a few seconds''}.

One developer said that \textit{``[...] where a new parameter is introduced it's hard to tell if the default value will do the trick, so I would likely have to do a decent amount of research before I felt secure in my choice [...]''}.

One developer identified a potential improvement when providing before/after examples rather than a complete migration, \textit{``When you provide before/after files for a migration it would be nice to have a quick summary of the migration, like did you add a parameter? Otherwise I have to look at the documentation to make sure I'm looking at the right thing, and takes a little bit of extra time.''}.

In one case a developer said that, \textit{``Small examples are nice, but sometimes you want more context, so maybe having the whole file is better''}, on the other hand another developer said that,\textit{''I prefer when I can only see the migration and a little surrounding code, if there's too much code, it's harder to see exactly what's going on.''}. From these interviews we believe that there is an element of developer preference when looking at examples, and perhaps future research can look at the kinds of examples developers like, and create tools that can determine how much of data-flow and control flow is shown based on developer preference.

\hypobox{
	\tool~can provide API migration assistance in 71/80 cases. We can automatically generate 14 faultless migrations, 21 migrations with minor code changes, and 36 migrations with useful guidance to developers. The effort needed to post-modify our generated API migration is low, an average of 3.65 tokens require modification.
	}

\noindent\textbf{RQ3: How much time can our approach save when migrating APIs?}
\label{sec:rq3}
\begin{table}[t]
	\vspace{-0.2cm}
	\centering
	\caption{Results of~\tool~user-study: comparing the time needed to migrate Android API usage examples (measured in seconds) with help from~\tool~and location-based API migration tools~\cite{dagenais_robillard_2009,wu_gueheneuc_antoniol_kim_2010}.}
	\vspace{-0.2cm}
	\label{tab:userStudy2}
	\begin{tabular}{c|rrrrr}
		Example & \begin{tabular}[c]{@{}r@{}}Avg.\\ time\\ w/o. A3\end{tabular} & \begin{tabular}[c]{@{}r@{}}Avg.\\ time\\ w. A3\end{tabular} & \begin{tabular}[c]{@{}r@{}}Time\\ Improvement\\ (\%)\end{tabular} & \multicolumn{1}{r}{\begin{tabular}[c]{@{}r@{}}Avg.\\ example\\ usefulness\end{tabular}} \\ \hline

		1     & 304.9 & 266.8 & 12.5 & 4.2 \\
		2     & 150.0   & 115.8 & 22.8 & 4.5 \\
		3     & 265.5 & 174.6 & 34.2 & 4.8 \\
		4     & 572.8 & 372.6 & 34.9 & 3.9 \\
		5     & 152.4 & 129.9 & 14.8 & 4.8 \\
		6     & 179.8 & 81.4 & 54.7 & 3.8 \\
		\hline
		Total: & 270.9 & 190.2 & 29.0 & 4.3 \\
	\end{tabular}%
	\vspace{-0.1cm}
\end{table}%

In this RQ, we present the design and results of a user study involving the assistance provided by our approach. We conduct the user study to evaluate the usefulness of migration suggestions provided by our approach. The user study involves 15 participants (6 professional developers and 9 graduate students). In particular, we compare the time used for API migration by using our approach, and by only using location-based API migration tools, such as SEMDIFF~\cite{dagenais_robillard_2009}, and AURA~\cite{wu_gueheneuc_antoniol_kim_2010}. We do not compare to LASE~\cite{Kim2013} in this section since LASE cannot identify migration cites or examples without developer input. Therefore, comparing A3, which automatically identifies a migration location, and automatically identifies potential migration examples, to LASE which can only migrate manually identified migration locations with manually identified examples, would not be a fair comparison of time saved. Rather, we seek to determine how much time can be saved by an approach that can automatically find migration examples and migration candidates when compared to approaches that only identify migration candidates.

We extracted six API migration tasks from the out-of-date Android API uses from FDroid projects. Each participant was given three tasks with the help of migration suggestions provided by~\tool, and three other examples with the help of location-based API migration tools, such as SEMDIFF~\cite{dagenais_robillard_2009}, and AURA~\cite{wu_gueheneuc_antoniol_kim_2010}. We randomized the order of the tasks for each participant. These tasks were part of the dataset used in RQ1 and RQ2 and had been detected by~\tool~as APIs which needed migration. The tasks were chosen randomly from our sample to avoid bias and we manually ensured that each task did need migration. We used the approach presented in Section~\ref{sec:step2_approach} to obtain a migration suggestion for each of the six tasks.

Each participant was given six source files that presented code with old versions of the Android API. The participants were told that they should attempt to modify the source code in each task to migrate to the latest version of the Android API. For the tasks that receive the help by~\tool, we provide the code that is generated by~\tool~after migration. For the tasks that only received the help from location-based API migration tools, the participants were informed of \emph{which} API call to migrate and \emph{what} is the new version of the API to migrate. We also provide the hyper-links to the Android developer website pages necessary to understand the API calls and their migrations were also given to the participants. We note that we did not directly run SEMDIFF~\cite{dagenais_robillard_2009} or AURA~\cite{wu_gueheneuc_antoniol_kim_2010} to obtain above information, but directly provide the ground truth information to the participants, as if SEMDIFF~\cite{dagenais_robillard_2009} or AURA~\cite{wu_gueheneuc_antoniol_kim_2010} generate perfect results. Furthermore, the participants were told that we provided potential solutions in the form of migration suggestions for some examples, and that they should attempt to use them if they could. This was done to minimize the noise in the measured time from other activities to concentrate on the migration of API calls themselves.

The participants were timed to determine how long each migration took. They were also asked to rank the usefulness of the migration examples whenever possible. The rank is on a scale of 1 to 5, where 1 was considered as useless, and 5 was considered extremely useful. The results of our user study are presented in Table~\ref{tab:userStudy2}.

Overall, our approach provides an average time improvement of 29\% with a p-value of 0.015 in a two-tailed Mann-Whitney U test. Professional developers improved by 45.9\% while graduate students by 22.7\%. We have therefore shown that automatically providing migration examples to users using a technique like~\tool~can improve migration times. As shown in Table~\ref{tab:userStudy2}, the developers involved with our user-study also found the assistance from~\tool~useful, ranking it an average of 4.3 out of 5 in usefulness. Therefore, our approach not only provides migration aid that can reduce migration time, but the examples provided are also judged as useful by developers.

\hypobox{\vspace{-0.05cm}Our approach provides, on average, a 29\% improvement in API migration speed compared with location-based API migration tools. Users ranked the help provided by~\tool~an average of 4.3 out of 5 on a usefulness scale.\vspace{-0.05cm}}
\vspace{-0.3cm}

\section{Related Work}
\label{sec:related}
In this section we discuss prior research in the field of APIs. We concentrate on prior work based on API evolution, API migration, and the usefulness of code examples for APIs. Table~\ref{tab:approach_differences} shows the major differences between our approach and prior studies. 

\begin{table*}[t]
	\vspace{-0.2cm}
	\caption{Main differences and novelty provided by the~\tool~approach when compared to related work.}
	\vspace{-0.3cm}
	\label{tab:approach_differences}
	\begin{tabular}{|cl|}
		\hline
		\multicolumn{1}{|c}{\textbf{Related Work}} & \multicolumn{1}{c|}{\textbf{Novelty provided by~\tool}} \\ \hline
		CatchUp!~\cite{Henkel2005} &  Our approach does not require recording of API modifications.\\
		SemDiff \& AURA~\cite{Dagenais2009, Wu2010} &  Our approach uses API usage examples rather than internal API modifications to obtain migration patterns.\\
		EXAMPLORE~\cite{Glassman_Zhang_Hartmann_Kim_2018} &  Our approach automatically obtains and applies relevant example code as patterns for migration.\\
		Sydit \& LASE~\cite{Meng_2011, Kim2013} &  Our approach automatically obtains matched code examples, does not require fully compilable projects.\\
		LibSync~\cite{Nguyen2010} &  Our approach gives fully migrated code, and does not require fully compilable projects.\\ \hline
	\end{tabular}
	\vspace{-0.4cm}
\end{table*}

\noindent\textbf{API evolution studies}

As the world of software APIs expanded so did the number of studies on API usage and evolution~\cite{Dig2005,Mendez_2013,Taneja_dig_xie_2007,businge_Serebrenik_brand_2013, wang_keivaloo_zou_2014,Li2013, Jaspan_2009, Dig_johnson_2006, Sridharan_2011, Wu_serveaux_2015, Holmes_Walker_2007, Businge2015, venkatesh_wang_zhang_zou_hassan_2016, Jaspan_Aldrich_2009, Godfrey_2001, deSouza_2004, Robillard_robert_2011}. More recently, several research papers have been published on the Android API and its evolution~\cite{lamothe_shang_2018,Li_Gao_2018,mcdonnell_ray_kim_2013,mehran_mahmoudi_sarah_nadi_2018,Linares_Vasquez_2013,Calciati_20189,Cai_2019,Li_bissyande_2016,LiLi_2018,li_gao_bissyand_ma_xia_klein_2020}. The Android API is both large in code size and has numerous versions. It has therefore been found to be a suitable case study for various types of software research, from API usage~\cite{wang_godfrey_2013} and crashes~\cite{Kong_2019}, to software testing~\cite{Moran_2018, Kong_Li_gao_2019}, Stack Overflow discussions~\cite{linares_vasquez_bavota_penta_oliveto_poshyvanyk_2014}, static analysis tool creation~\cite{LiLi_Bis_2019}, and user ratings~\cite{Bavota2015}. We use the Android API as a case study for API evolution and migration.

In 2013, McDonnell et al.~\cite{mcdonnell_ray_kim_2013} extracted the change information of the Android API. They found that the Android API is rapidly evolving and provides an average of 115 API updates per month. API updates are said to occur to fix bugs, enhance performance, and to respect new standards~\cite{mcdonnell_ray_kim_2013}. They suggest that although there are currently many tools to automate the API updating process, these tools are insufficient for current needs and that new studies and tools are necessary to encourage proper API updates~\cite{mcdonnell_ray_kim_2013}. The work by McDonnel et al. is a strong motivation for the ideas presented in this paper. Due to the rapid nature of changes in the Android API, it is necessary to have approaches such as the one presented in this paper to aid developers in adapting API changes. %to changes in an API.

Li et al.~\cite{Li_Gao_2018,li_gao_bissyand_ma_xia_klein_2020} built a tool to investigate deprecated Android APIs. Li et al. find that 37.87\% of apps make use of deprecated APIs~\cite{Li_Gao_2018}. This knowledge further strengthens the need for tools to help developers migrate away from these methods. They also found that although the Android framework developers consistently document replacements for deprecated APIs, this documentation is not always up to date~\cite{Li_Gao_2018}. This presents a perfect opportunity for our approach to use the most up-to-date to help users migrate away from deprecated API rather than rely on potentially outdated documentation.

\noindent\textbf{API migration techniques}

Past research has presented tools and suggestions to improve API migrations~\cite{henkel_diwan_2005,dagenais_robillard_2009,wu_gueheneuc_antoniol_kim_2010,phan_nguyen_et_al_2017,Langdon_White_2016,Li_Wang_Yingfei_hu_2015,wang_many_to_many_map,Nguyen2016,Nguyen_Nguyen_Nguyen_2014,Nguyen_trong_duc_nguyen_2016,zhong_tao_lu_jian_hong_2009, phan_nguyen_et_al_2017}. API migration tools usually rely on fully automatic or semi-automatic means to provide API migration recommendations to developers. These approaches concentrate on locating a migration candidate and recommending an alternative API. Meanwhile, our approach assumes that the user knows that an alternative API exists, either through these tools or through documentation, and does not understand \textit{how} to produce the migration. This is where our approach,~\tool, aims to guide users.

\textit{CatchUp!}~\cite{henkel_diwan_2005} provides a semi-automatic migration solution which requires that API developers record API source code changes from their IDE. These changes can then be replayed on API user systems to adapt their application to the changed API. This tool can provide safe migrations due to the nature of the tool chain. However, the tool requires the API developers to record their own work. Contrarily to \textit{CatchUp!},~\tool~uses a myriad of examples to guide the migration from any available source.

Dagenais and Robillard, and Wu et al. provide different semi-automatic migration tools, \textit{SemDiff}~\cite{dagenais_robillard_2009}, and \textit{AURA}~\cite{wu_gueheneuc_antoniol_kim_2010}. Both tools similarly mine method changes in two versions of an API, and internal adaptations to these API changes. These changes are then used to provide migration recommendations to API users. \textit{SemDiff} presents migration scenarios by heuristically ranking the most likely candidates. Meanwhile, \textit{AURA} uses heuristics to provide only the best fit migration.~\tool~can consider any source of example to extract migrations. This allows users to obtain high quality migration examples even if the API source code is proprietary, as long as open source examples of migrations exist or can be created.

Similarly to~\tool, more recent approaches such as Meditor~\cite{Xu_2019} and the work by Fazzini et al.~\cite{Mattia_2019}, attempt to mine repositories to obtain examples and apply those examples to real world applications. Meditor~\cite{Xu_2019} concentrates on Java applications and manages to correctly apply code migrations in 96.9\% of cases. Other similar approaches~\cite{Mattia_2019} have since been attempted on the Android ecosystem with a success rate of 85\%. However, to the best of our knowledge, our approach is the only one so far to have included evaluation in real world scenarios with a user study and developer interviews.

\noindent\textbf{Code examples and APIs}

~\tool~relies on the existence of source code examples of API migrations. Prior research has covered API documentation enhancement and API examples extensively~\cite{Nguyen_rigby_2016, Buse_Weimer_2012, Mandelin_Xu_Kimelman_2005, Wang_dang_zhang_chen_xie_zhang_2013,Meng_2013, Niu_Keivanloo_Zou_2017, Phan_Ng_truong_2018, Moritz2013, Subramanian_2014, Hu_Li_Xia_Lo_Lu_Jin_2018, Zhai_2016, parnin_treude_2011, Zhong_zhendong_2013, Wang_Godfrey_2015}.

Gao et al. find that crypto-APIs misuses are very common in Android apps~\cite{Gao_2019}. They find that developers are unlikely to successfully fix these misuses when they attempt to do so~\cite{Gao_2019}. These findings highlight the importance of proper testing and keeping the developer in the development loop. This is one of the reasons why A3 is not a fully automated approach, but rather allows the developer to have the final say in any migration decision. Therefore, it is always possible for the developer to determine the validity of any migration before choosing to use it.

\textit{EXAMPLORE} is a visualization tool to assist users in understanding common uses of APIs~\cite{Glassman_Zhang_Hartmann_Kim_2018}. The tool uses a corpus of API examples to build common usage patterns which can then be displayed to a user in an interactive graphical way. Since tools like \textit{EXAMPLORE} have shown that they can assist users to answer API questions with detail and confidence~\cite{Glassman_Zhang_Hartmann_Kim_2018}, we believe that using examples to provide API insights could similarly be used for a migration tool.

Tools such as \textit{Sydit}, \textit{LibSync}, and \textit{LASE} provide systematic editing through examples~\cite{Meng_2011,Nguyen2010,Meng_2013}. However, these tools require well-built ASTs for dependency analysis, but examples of API migration may not always be compilable (e.g., code snippets from documentation, or examples from users). Furthermore, a developer has to provide the examples with relevant context, or in the case of \textit{LibSync} project versions where migrations are known to occur (e.g., where to apply the changes). In contrast, our approach automatically mines relevant examples, and then finds the most relevant migration pattern for a specific API migration use case with no dependency assumptions (i.e., the code can be non-compilable).

\textit{MAPO} is a tool that allows developers to rapidly search for frequent API usages~\cite{Xie_pei_2006}. MAPO uses a combination lightweight source code analysis, code search engines, and frequent item-set mining methods~\cite{Xie_pei_2006}. Tools like \textit{MAPO} have shown that it is possible to rapidly and reliably find examples of API usage on the web. Therefore, tools like~\tool~can rely on this foundation to use these examples for more complex tasks, such as aiding API migration.

We do not provide direct comparisons with other approaches as the work presented in this paper does not directly compare to previous approaches but rather augments them. It would be possible to use the example gathering power provided by~\tool~and the example matching approaches provided by Sydit, LASE, or LibSync to complete the migration process with minor modifications to these approaches. However, in our partial and indirect comparison with LASE, SEMDIFF~\cite{dagenais_robillard_2009}, and AURA~\cite{wu_gueheneuc_antoniol_kim_2010} we find that our approach matches or outperforms prior approaches. The primary novelty of our approach is getting migration patterns directly from publicly available sources without needing the compilation of the sources, or developer intervention, and directly using these sources to determine if a migration pattern can be used to modify developer code, and then automatically attempting the migration and providing this to the user. No other approach to date covers the full scope of our approach and therefore no direct and fair comparison can be made.

\section{Threat to Validity}
\label{sec:threats}
\noindent\textbf{\textit{Construct validity.}}
We assess the validity of our API migrations by building and running the apps as well as running the test suites of the migrated Android apps. Although we focus on the tests that exercise the migrated API calls, and attempt to exercise as much functionality as possible when running the apps, it is still possible that defects introduced by the migration are not identified by our tests. User studies and interviews with developers may complement the evaluation of our approach. In our study, there are still cases where our approach cannot migrate faultlessly. Although our approach can provide migration guidelines to developers, as an early attempt of this line of research, our approach can be further complemented by other techniques such as code completion to achieve better assistance in API migration. Furthermore, we concentrate on the majority of cases through file level migrations. If a large number of migrations occur across multiple files, our approach is not currently able to help.

\noindent\textbf{\textit{External validity.}}
Since this entire study was tested on the Java API of the Android ecosystem, it is possible that the findings in this paper will not generalize to other programming languages. However, while it is true that the approach presented in this paper was tested specifically on a Java based API, all of the approaches are built upon assumptions that are true in other popular programming languages such as C\#.

\noindent\textbf{\textit{Internal validity.}}
Our findings are based on the Android project and code examples mined and produced for its API. It is possible that we only found a subset of all migrations. It is also possible for the time gap between the release of new API and the update to examples to be larger in other sources. We attempted to mitigate these threats through mining official samples, open source projects, and having participants produce examples for frequently used APIs. We found that Google Samples updated deprecated API as soon as one month after the release of a new API version, which should allow developers to regularly update their apps. Our participant created examples were new and useful API migration examples, showing that the premise of using examples to help automate API migrations is functional and likely dependent on the sample size of examples.

\section{Conclusion}
\label{sec:conclusion}
In this paper, we proposed an approach that assists developers with Android API migrations by learning API migration patterns from code examples mined directly from available code repositories. We evaluate our approach by applying automated API migrations to 32 open-source Android apps from FDroid and through a user-study. We find that our approach can automatically extract API migration patterns from both public code example and manually produced API examples that are created with minimal effort. By learning API migration patterns from these examples, our approach can provide either automatically generated API migrations or provide useful information to guide the migrations. Our user-study showed that the examples provided by our approach allow users to migrate Android APIs, on average, 29\% faster and are seen as useful by developers, who ranked them an average usefulness of 4.3 out of 5.

This paper makes the following contributions:
\begin{itemize}
	\item We propose a novel approach that learns API migration patterns from code examples taken directly from available code repositories.
	\item Our novel approach can automatically assist in API migration based on the learned API migration patterns. 
	\item We produce a user study and conduct semi-structured interviews that conclusively shows that migration examples are both desirable and useful to developers.
\end{itemize}

Our approach illustrates the rich and valuable information in code examples that can be leveraged in API related software engineering tasks.

\ifCLASSOPTIONcaptionsoff
\newpage
\fi

\bibliographystyle{IEEEtran}
\bibliography{bibliography}

\begin{IEEEbiography}[{\includegraphics[width=1in,height=1.25in,clip,keepaspectratio]{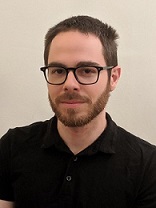}}]{Maxime Lamothe}
	Maxime Lamothe is a Ph.D. student in the department of Computer Science and Software Engineering at Concordia University, Montreal, Canada. He has received his M.Eng. degree from Concordia University and he obtained his B.Eng. from McGill University. His research interests include API evolution, API migration, machine learning in software engineering, big data software engineering, and empirical software engineering.
\end{IEEEbiography}

\begin{IEEEbiography}[{\includegraphics[width=1in,height=1.25in,clip,keepaspectratio]{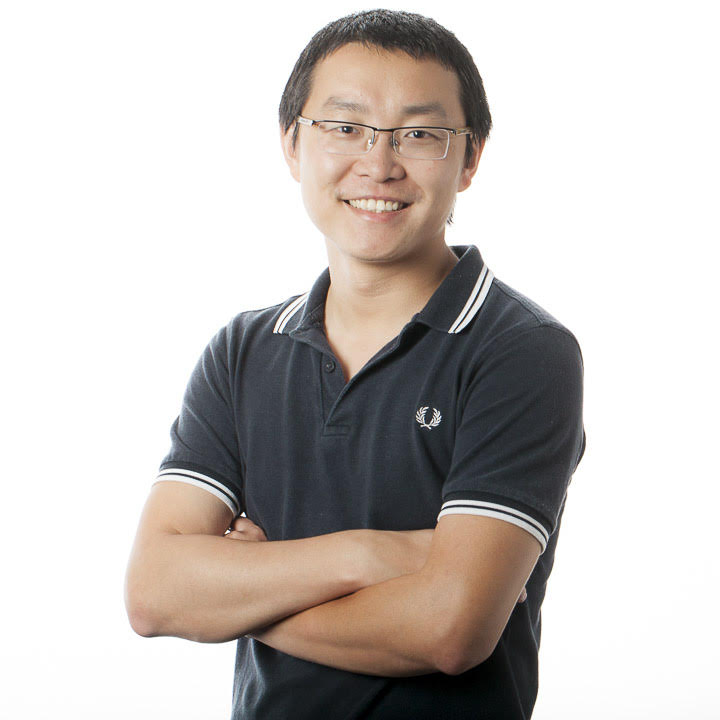}}]{Weiyi Shang}
	Weiyi Shang is a Concordia University Research Chair at the Department of Computer Science. His research interests include AIOps, big bata software engineering, software log analytics and software performance engineering. He is a recipient of various premium awards, including the SIGSOFT Distinguished paper award at ICSE 2013, best paper award at WCRE 2011 and the Distinguished reviewer award for the Empirical Software Engineering journal. His research has been adopted by industrial collaborators (e.g., BlackBerry and Ericsson) to improve the quality and performance of their software systems that are used by millions of users worldwide. Contact him at shang@encs.concordia.ca; http://users.encs.concordia.ca/\texttildelow shang.
\end{IEEEbiography}

\begin{IEEEbiography}[{\includegraphics[width=1in,height=1.25in,clip,keepaspectratio]{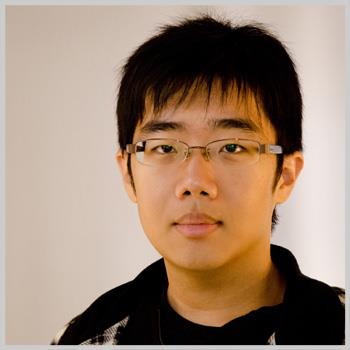}}]{Tse-Hsun~(Peter)~Chen}
	Tse-Hsun (Peter) Chen is an Assistant Professor in the Department of Computer Science and Software Engineering at Concordia University, Montreal, Canada. He leads the Software PErformance, Analysis, and Reliability (SPEAR) Lab, which focuses on conducting research on performance engineering, program analysis, log analysis, production debugging, and mining software repositories. His work has been published in flagship conferences and journals such as ICSE, FSE, TSE, EMSE, and MSR. He serves regularly as a program committee member of international conferences in the field of software engineering, such as ASE, ICSME, SANER, and ICPC, and he is a regular reviewer for software engineering journals such as JSS, EMSE, and TSE. Dr. Chen obtained his BSc from the University of British Columbia, and MSc and PhD from Queen's University. Besides his academic career, Dr. Chen also worked as a software performance engineer at BlackBerry for over four years. Early tools developed by Dr. Chen were integrated into industrial practice for ensuring the quality of large-scale enterprise systems. More information at: http://petertsehsun.github.io/.
\end{IEEEbiography}
\clearpage

\end{document}